# Diversity and network coherence as indicators of interdisciplinarity: Case studies in bionanoscience

Ismael Rafols
SPRU, University of Sussex, Brighton, UK
i.rafols@sussex.ac.uk

Martin Meyer
SPRU, University of Sussex, Brighton, UK
Steunpunt O&O Statistieken, Katholieke Universiteit Leuven, Belgium
Helsinki University of Technology, Finland

June 30th, 2008

**Abstract**

The multidimensional character and inherent conflict with categorisation of interdisciplinarity makes its mapping and evaluation a challenging task. We propose a conceptual framework that aims to capture interdisciplinarity in the wider sense of knowledge integration, by exploring the concepts of diversity and coherence. Disciplinary diversity indicators are developed to describe the heterogeneity of a bibliometric set viewed from predefined categories, i.e. using a top-down approach that locates the set on the global map of science. Network coherence indicators are constructed to measure the intensity of similarity relations within a bibliometric set, i.e. using a bottom-up approach, which reveals the structural consistency of the publications network. We carry out case studies on individual articles in bionanoscience to illustrate how these two perspectives identify different aspects of interdisciplinarity: disciplinary diversity indicates the large-scale breadth of the knowledge base of a publication; network coherence reflects the novelty of its knowledge integration. We suggest that the combination of these two approaches may be useful for comparative studies of emergent scientific and technological fields, where new and controversial categorisations are accompanied by equally contested claims of novelty and interdisciplinarity.

**Keywords**
Interdisciplinary research; nanotechnology; nanoscience; diversity; indicators; network analysis.

## 1. Introduction

In policy discourse interdisciplinarity is often perceived as a mark of 'good' research: interdisciplinary research is seen as more successful in achieving breakthroughs and relevant outcomes, be it in terms of innovation for economic growth or for social needs. This has led to policies aimed at fostering interdisciplinarity, particularly in fields, such as biotechnologies or nanotechnologies, regarded as emerging through technological convergence.

However, there is no systematic evidence, to our knowledge, showing that "more" interdisciplinarity leads to "better" research, although there is plenty of anecdotal evidence suggesting that interdisciplinarity has played a crucial role in many scientific breakthroughs (e.g. Hollingsworth's (2006) work on biomedical research). This lack of evidence stems from the difficulties of agreeing on criteria to define scientific performance (a complex issue we will not discuss here) and intensity of interdisciplinarity (the aim of this paper).

The concept of interdisciplinarity and its variants (multi, trans, crossdisciplinarity)[1] is problematic, if not controversial (Weingart and Stehr, 2000). First, given its polysemous and multidimensional nature (Sanz-Menéndez et al., 2001), there is no agreement about pertinent indicators, or the appropriateness of categorisation methods (Bordons et al., 2004). Second, although the etymology of inter-, multi-, trans- and cross-disciplinarity suggests that this is a property that is between, beyond or across various disciplines, interdisciplinarity is widely and ambiguously used to mean research spanning a variety of areas - academic disciplines, technological fields and/or even industrial sectors. Consequently, interdisciplinarity has been declared to be 'no longer adequate' (Klein 2000, p.3) or a misnomer (Glaser, 2006). Thus, the process of integrating different bodies of knowledge rather than transgression of disciplinary boundaries *per se*, has been identified as the key aspect of so-called 'interdisciplinary research' (National Academies, 2005).

How can this knowledge integration be assessed? While some sort of taxonomy is necessary to 'shrink' and locate on a manageable map the integration occurring in the gigantic landscapes of scientific knowledge, any categorisation entails the adoption of 'rigid' boundaries, which hinders accurate description of the 'fluid' dynamics of science (Zitt, 2005). This tension between taxonomy and dynamics is particularly acute in emergent fields, and often produces conflicting views. For example, in nanotechnology, coarse-grained studies tend to emphasize the interdisciplinary nature of the field (Meyer and Persson, 1998; Leydesdorff and Zhou, 2007), whereas lower level studies suggest that, below the re-labelling, genuine knowledge integration is occurring at a slower pace (Schummer, 2004; Rafols, 2007).

Policies fostering interdisciplinarity, therefore, sometimes appear to be based more on conventional wisdom and arbitrary classification than on empirical evidence. This investigation aims to inform policy-making on the dynamics of emerging fields by providing measures that capture the intensity of interdisciplinarity in the wider sense of knowledge integration. We do so by combining macro and micro level perspectives. We use as case studies individual publications in biomolecular motors, a research specialty of bionanoscience, and analyse interdisciplinarity as revealed from the set of references.

The paper is organised as follows. Section 2 presents the concepts of diversity and coherence, relates them to the literature on interdisciplinarity and proposes an analytical framework to investigate knowledge integration. Section 3 describes the empirical data and the operationalisation of diversity and coherence as bibliometric indicators. Section 4 applies the diversity-coherence framework to case studies of individual articles in biomolecular motors. Section 5 summarises the results and discusses their implications.

---

[1] In this study interdisciplinarity refers to all these types of cross-disciplinary research.

## 2. Conceptual framework

### 2.1. Review of bibliometric studies on interdisciplinarity

Several bibliometric studies have addressed the issue of interdisciplinarity, directly (see review by Bordons et al., 2004) or through discussion of related issues such as mapping knowledge flows among fields (see review by Zitt, 2005). Its study involves the choice of a disciplinary taxonomy, and/or relational properties (similarities, co-occurrences, flows) to characterise the interactions between elements or categories.

Most investigations use a top-down approach and predefined categories (typically ISI Subject Categories - SCs) to study their proportions and/or relations. For example, van Raan and van Leeuwen (2002) describe interdisciplinarity in an institute in terms of the percentage of publications and citations received to and from each SCs. In the following three sections, we explore how these studies can be conceptualised as expressing *disciplinary diversity*.

Some investigations adopt a bottom-up approach, in which the low-level elements investigated (e.g. publications, papers) are clustered or classified into factors on the basis of multivariate analyses of similarity measures (Small, 1973; Braam et al., 1991; van den Besselaar and Leydesdorff, 1994; Schmidt et al., 2006). These clusters are then projected in 2D or 3D maps to provide an insight into the structure of the field and estimate the degree of network-level similarity. Similarity measures have also been used to compute network properties, such as centralities, to identify interdisciplinarity (Otte and Rousseau, 2002; Leydesdorff, 2007). Following Nesta and Saviotti (2005), in this study, we conceptualise network-level properties as *network coherence*.

We build on top-down and bottom-up approaches, to develop a methodology combining (i) diversity measures using large-scale disciplinary categories, with (ii) network measures based on similarities among publications.

### 2.2. Definition of interdisciplinarity

In line with a number of works (National Academies, 2004, Porter et al., 2006, p.3), interdisciplinarity is defined here as a mode of research that integrates concepts or theories, tools or techniques, information or data from different bodies of knowledge. As highlighted by Porter et al., the key concept is 'knowledge integration'. In order to capture the process of integration in research, we need to investigate two aspects:

> **Diversity:** number, balance and degree of difference between the bodies of knowledge concerned;
>
> **Coherence:** the extent that specific topics, concepts, tools, data, etc. used in a research process are related.

In this framework, we view the knowledge integration process as being characterised by high cognitive heterogeneity (diversity) and increases in relational structure (coherence); in other words as a process in which previously different and disconnected bodies of research become related.

### 2.3. Diversity: concept and measures

The concept of diversity is used in many scientific fields, from ecology to economics and cultural studies, to refer to three different attributes of a system comprising different categories (Stirling, 1998, 2007; Purvis and Hector, 2000):

- **variety**: number of distinctive categories;
- **balance**: evenness of the distribution of categories;
- **disparity or similarity** [2] **:** degree to which the categories are different/similar.

Figure 1 depicts how an increase in any of these attributes results in an increase in the diversity of the system examined.

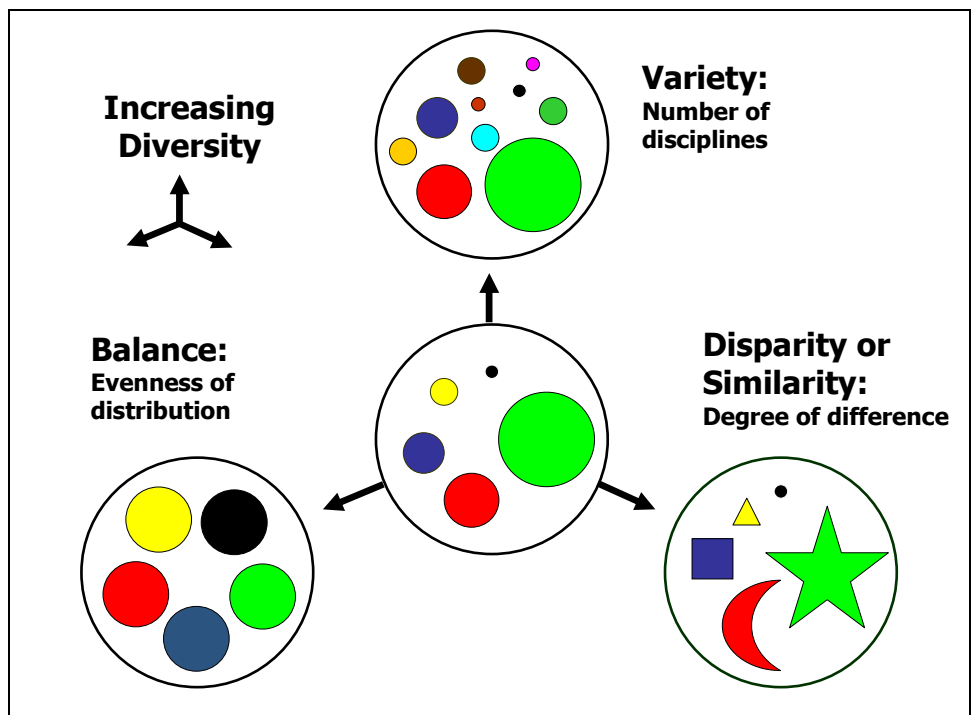

*Figure 1. Schematic representation of the attributes of diversity, based on Stirling (1998, p. 41).*

---

[2] Hereafter we will use only the term similarity, which is the one commonly used in bibliometrics.

Stirling (2007) shows that classic indices of diversity, such as Shannon's or Simpson's (also known as Herfindahl's), measure a combination of variety and balance, but fail to account for the distances or similarities between categories. On the basis of a set of criteria, he proposes a general diversity heuristic in order to explore how diversity indices differ when more or less emphasis is given to variety, balance and similarity. Stirling's heuristic can be formulated in a generalised diversity index which reduces to the traditional indices for specific set of parameters α, β (Stirling, 2007, p.7). For parsimony, here we define Stirling index Δ as the variant for α=1, β=1, the simplest form incorporating variety, balance and similarity. Table 1 presents the notation and diversity indices used in our study.

Table 1. Selected measures of diversity.

| Notation: | |
|---|---|
| Proportion of elements in category $i$: | $p_i$ |
| Distance between categories $i$ and $j$: | $d_{ij}$ |
| Similarity between categories $i$ and $j$: | $s_{ij} = 1 - d_{ij}$ |
| **Indices:** | |
| $N$ = Variety | $N$ |
| $H$ = Shannon | $-\sum_i p_i \ln p_i$ |
| $I$ = Simpson diversity[3] | $\sum_{i,j(i \neq j)} p_i p_j = 1 - \sum_i p_i^2$ |
| $\Delta$ = Stirling (α=1, β=1) | $\sum_{i,j} d_{ij} p_i p_j = 1 - \sum_{i,j} s_{ij} p_i p_j$ |
| Generalised Stirling | $\sum_{i,j} d_{ij}^\alpha (p_i p_j)^\beta$ |

As the formulae show, Stirling index Δ can be understood as a Simpson diversity in which the products of proportions of categories are weighted by distance/similarity. Our interest in using Stirling's framework to track interdisciplinarity is twofold. First, since Stirling's generalised formulation needs a metric ($d_{ij}$) and has open values for the parameters α and β, it highlights that the mathematical form of any diversity index includes some prejudgement of the aspect of diversity that is considered important. High values for β give more weight to the contribution of large categories, and high values for α see the co-occurrence of distant categories as more important. The choice of the metric used to define distance is inevitably value laden. Second, and very importantly for emerging fields, the inclusion of distance among categories lessens the effect of inappropriate categorisation changes: if a new category $i$ is very similar to an existing category $j$, their distance $d_{ij}$ will be close to zero, and its inclusion in categories list will result in only slightly increased diversity.[4]

---

[3] Simpson diversity is defined as (1-Simpson Index), where the Simpson index is the commonly used measure of concentration.
[4] One example could be 'Nanoscience&Nanotechnology' (N&N) from the ISI categorisation: according to Leydesdorff's and Rafols' metric (submitted), N&N has a distance of only 0.0354

In the next section, we explore how these measures relate to already developed measures of interdisciplinarity.

*2.4. Use of diversity in studies of interdisciplinarity*

In this section, we present some illustrations of how bibliometric studies explicitly or implicitly address the properties of diversity, namely variety, balance and similarity, when investigating interdisciplinarity:

**Variety:** Morillo et al. (2003, p. 1241) for each SC, counted the number of other SCs with which it shared journals. Presentations of disciplinary profiles, e.g. in bar charts, provide visual cues for this variety (e.g. van Raan and van Leeuwen, 2001, p. 611).

**Balance:** Since Porter and Chubin's (1985) seminal contribution, perhaps the most common indicator of interdisciplinarity has been the percentage of citations outside the discipline of the citing paper. Van Leeuwen and Tijssen (2000) showed that this could be as high as 69% on average. Similarly, Schummer (2004, p. 449) uses the percentage of co-occurrences of affiliations based on different disciplines as indicator.

**Similarity:** Measures of similarity among predefined categories have been widely used to visualise the relative positions of different scientific disciplines (Moya-Anegón et al., 2004, 2007). Although in most cases associated dissimilarity values are not presented, the visualisation implicitly conveys the degree of diversity.[5]

In some instances, these three properties are explicitly addressed in the same study. An interesting case is Morillo et al.'s work on the multi-assignation of journals to SCs (Morillo et al., 2003; Bordons et al., 2004, pp. 447-453). For each category, these studies looked at:

- the *balance*, in terms of percentage of multi-assigned journals for one SC;
- the *variety* of links with other SCs, namely the number of different SCs with which a given SC shares journals;
- the strength of linkages (or *similarities*) given by number of co-assigned journals for two SCs.

This multidimensional approach, covering different aspects of disciplinary diversity, allowed Morillo and co-workers to develop an elaborate taxonomy of interdisciplinarity types across science fields.

---

with 'Materials Science, multidisciplinary', whereas the distance between the latter and a relatively related field, such as 'Physics, applied', is 0.1916.

[5] Matrices of knowledge flows among disciplines are another way to present interdisciplinarity. E.g. Bourke and Butler (1998), calculated the number of publications from discipline-based departments associated to discipline-based journals. These matrices can then be used to compute similarity measures.

Finally, some studies use more complex indicators, such as the Pratt number (similar to Simpson's; see Morillo, 2001), or Shannon entropy (Barjak, 2006), which combine the properties of variety and balance.[6]

While the bibliometric studies referred to above touch on particular aspects of diversity, to our knowledge, only the recent paper by Porter et al. (2007) actually integrates the attributes of variety, balance and similarity into one index. Interestingly, Porter's indicator of *Integration* is a particular parameterisation of Stirling's index Δ (see Table 1), where the similarities $s_{ij}$ are Salton's cosines for co-citation patterns among ISI SCs. Here, we operationalise Stirling's diversity following Porter's indicator, as described in the Data and Methods section.

### 2.5. Coherence: concept and measures

The concept of coherence aims to capture the extent to which of a system's elements are consistently articulated and form a meaningful constellation (Stirling, personal communication). Hence, coherence is a general property that addresses the functionality of a system. In our bibliometric context, coherence expresses the extent to which publication networks form a more or less compact structure. If we take degree of cognitive similarity as the linkage between publications (e.g. by using co-citation, co-word or bibliographic coupling), a more clustered network is seen as having higher cognitive coherence.

Coherence, or cognate concepts such as cohesion or compactness, have been extensively investigated in information sciences (see Egghe and Rousseau, 2003, for a bibliometric discussion). In the context of economic studies of innovation, coherence has been utilised to account for the aggregated relatedness (or similarity) of the firm's technological base, with the idea that "coherent firms are more likely to be successful than incoherent ones" (Nesta and Saviotti, 2005, p.124). Here, we introduce coherence in order to express the degree of integration already in place in a body of research. However, since the key aspect of interdisciplinary research has been argued to be the *process* of knowledge integration (Section 2.2), interdisciplinarity should ideally be assessed in terms of a temporal derivative, i.e. a change in coherence.

Depending on the unit of analysis used in the study of interdisciplinarity, coherence can take different meanings. High coherence within the reference set in a publication means that its referencing practices are highly specialised and hence, that it builds on an already established research specialty. High coherence in the publication set of an interdisciplinary centre would suggest that it is achieving its integrative mission.

---

[6] Other publications use measures of diversity in bibliometrics, to examine not the diversity of disciplines, but diversity/concentration of research in institutions (e.g. Rousseau, 2000).

Since our definition of coherence is in terms of the network of relations among the basic elements (publications), it has to be operationalised using bottom-up approaches. This avoids the use of previous categorisations but requires spiralling computing efforts for large data sets.

In our view, bibliometric studies related to network coherence fall into the areas of mapping and associated methods of clustering, along with other multivariate analyses based on low level categories such as single articles, authors or journals. An example is Small's (1977) study of a research specialty over five years. Using co-citation analysis, Small tracked the appearance and disappearance of clusters in the research specialty and proposed a 'Stability Index', based on degree of overlap between the clusters, that described the coherence of the network. Other examples are combinations of co-citation and co-word analysis (Braam et al., 1991), and large scale mapping using inter-citation flows among journals (Boyack et al., 2005). Methodologies from network analysis continue to be experimented with, as shown by Hellsten et al.'s (2007) adoption of an Optimal Percolation Method, and Schmidt et al.'s (2006) clustering of research fronts. The connection between interdisciplinarity and network structure, as shown by factor analysis, was made explicit by van den Besselaar and Heimeriks (2001). More recently Leydesdorff (2007) explored network centralities as indicators of interdisciplinarity. Building on these network approaches, we use simple network analysis measures for the operationalisation of coherence, as described in the Data and Methods section.

*2.6. Disciplinary diversity vs network coherence*

We introduced the concepts of diversity and coherence in relation to interdisciplinarity and have shown how they are related to previous bibliometric investigations. Table 2 presents a summary of this conceptual framework. In this subsection we argue for the need to combine disciplinary diversity and network coherence analyses to achieve a more nuanced view of knowledge integration.

As discussed above, the problem with disciplinary diversity is that it relies on predefined and 'rigid' categories, which may miss emergent or dynamic phenomena in science. The inclusion of metrics between categories (as in Stirling's index) lessens the effect of creating very similar categories, but does not solve the problem of hidden divides within existing categories.

Coherence approaches might be seen as being more accurate, but unfortunately they present a very problematic trade-off between size and level of analysis. For micro- or meso-level investigations, bottom-up network approaches are more accurate for describing direct knowledge flows or other explicit relations. However, they cannot capture the position of local elements in the global map of science, and thus miss the large-scale perspective of the integration process. At the other extreme, in macro-level studies using complicated metrics, the direct relations between elements become opaque. In addition, the use of large bibliometric sets requires access to expensive databases and computational resources that are beyond the reach of most researchers.

Table 2. Summary of conceptual framework.

|  | **Diversity** | **Coherence** |
|---|---|---|
| General concept: | Heterogeneity in terms of variety, balance and similarity of categories | Functional articulation and structural compactness of elements in system |
| Main research tradition: | Ecology | Network Analysis |
| Type of approach: | Top-down | Bottom-up |
| Categorisation: | Pre-defined | Unnecessary |
| Metric: | Optional (needed for Stirling) | Necessary |
| Indices: | $N$ = Variety<br>$H$ = Shannon<br>$I$ = Simpson<br>$\Delta$ = Stirling | $S$ = Mean Linkage Strength<br>$L$ = Mean Path Length |

Given these constraints, we propose to combine disciplinary diversity (top-down) and network coherence (bottom-up) perspectives to track knowledge integration in small and medium sized studies. Figure 2 provides a schematic representation of this twofold perspective, after Porter et al.'s (2007, p. 139) proposal. Since in this study we take individual publications and study knowledge integration through their reference sets, each of the nodes in the networks represents a reference, and each link the degree of similarity between references (we use bibliographic coupling). There are four possible combinations:

(i) **Low** diversity - **High** coherence is a case of specialised disciplinary research –all the references are from the same discipline and are related.
(ii) **Low** diversity - **Low** coherence is a case of a publication relating distant research specialties within one discipline.
(iii) **High** diversity - **Low** coherence is a case of a publication citing references that were hitherto unrelated and belong to different disciplines: a potential instance of interdisciplinary knowledge integration.
(iv) **High** diversity - **High** coherence is a case of a publication citing across several disciplines, to references that are similar. This similarity suggests that the references belong a single research specialty. Hence, although the publication is interdisciplinary, it does not involve new knowledge integration.

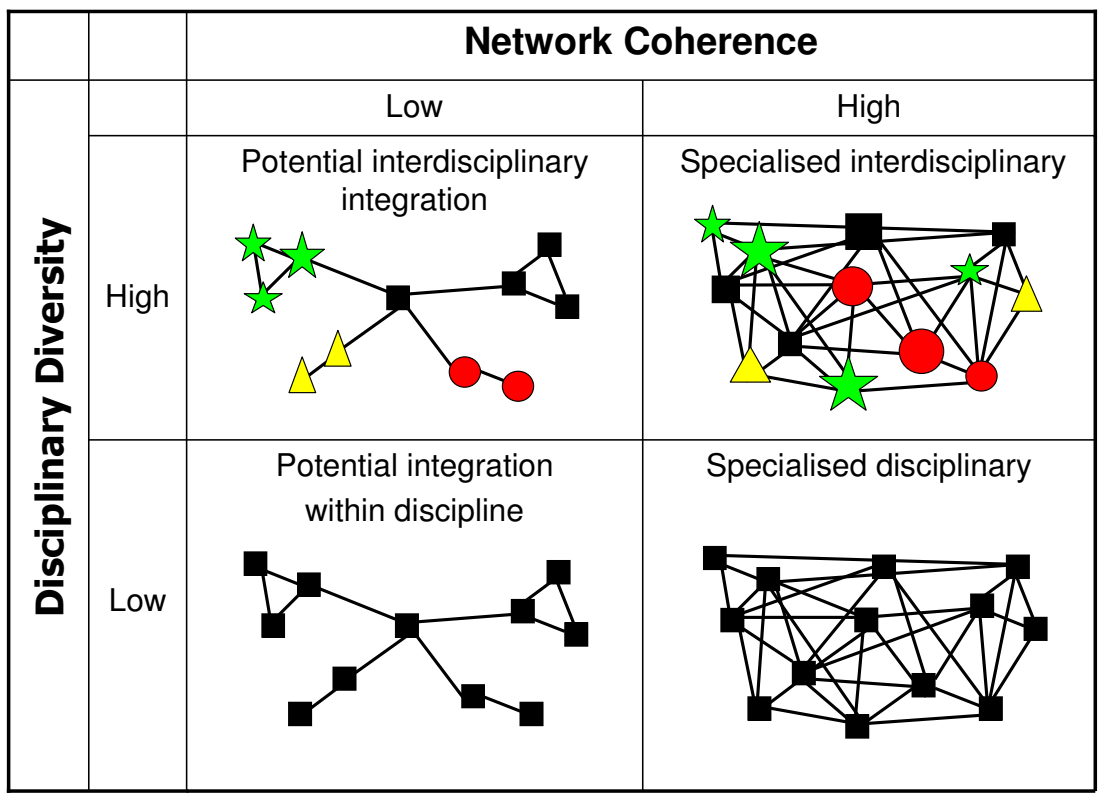

*Figure 2. Disciplinary diversity vs network coherence.*

Figure 2 provides a simple heuristics to trace knowledge integration. However, as discussed above, although low coherence suggests potential integration, we would need to examine the process, i.e. the trajectory over the matrix, to confirm this. Knowledge integration trajectories should move from left to right, from less to more coherence.

This scheme is partly based on Porter et al.'s (2007) framework. They rely on the combination of two indicators based on ISI SCs: *Integration* and *Specialisation*. Integration captures the diversity of SCs in the references of the set of papers; specialisation is the reverse of diversity (i.e. 1-$\Delta$, in its most recent formulation) for the SCs of the journals in which the papers are published. The distinction between diversity in referencing and publishing is insightful and useful to differentiate between multidisciplinary and interdisciplinary research. However, since both integration and specialisation are based on ISI SCs, they are correlated. The complement of network coherence is useful; since its indicators are based on data and methods independent of SCs, they contribute an 'orthogonal' perspective on knowledge integration.

## 3. Data and methods

*3.1. Data*

This study builds on previous investigations of interdisciplinary practices in laboratories of biomolecular motors, one of the specialties in bionanoscience (Rafols and Meyer, 2007). From the keynote speakers at a 2005 international conference on biomolecular motors, we selected the Japanese researchers and interviewed them about a specific project, as perceived in the light of two or three major publications. We inquired into their affiliations, backgrounds, the techniques and instruments used and how they were acquired, their collaborations, and the story of the research process. These data were complemented by information from scientific publications, miscellaneous documentation and homepages. Detailed data for these case studies was presented in Rafols and Meyer (2007) and discussed in Rafols (2007).

From the ISI Web of Science we downloaded full bibliometric records for the publications on which we had based our interviews. These records were processed using the bibliometric programme Bibexcel (Persson, 2008), the statistical packet R (2007), and the network analysis software Pajek (Batagelj and Mvar, 2008). For each publication, diversity and coherence measures were computed as summarised in Figure 3 and described in the following two subsections.

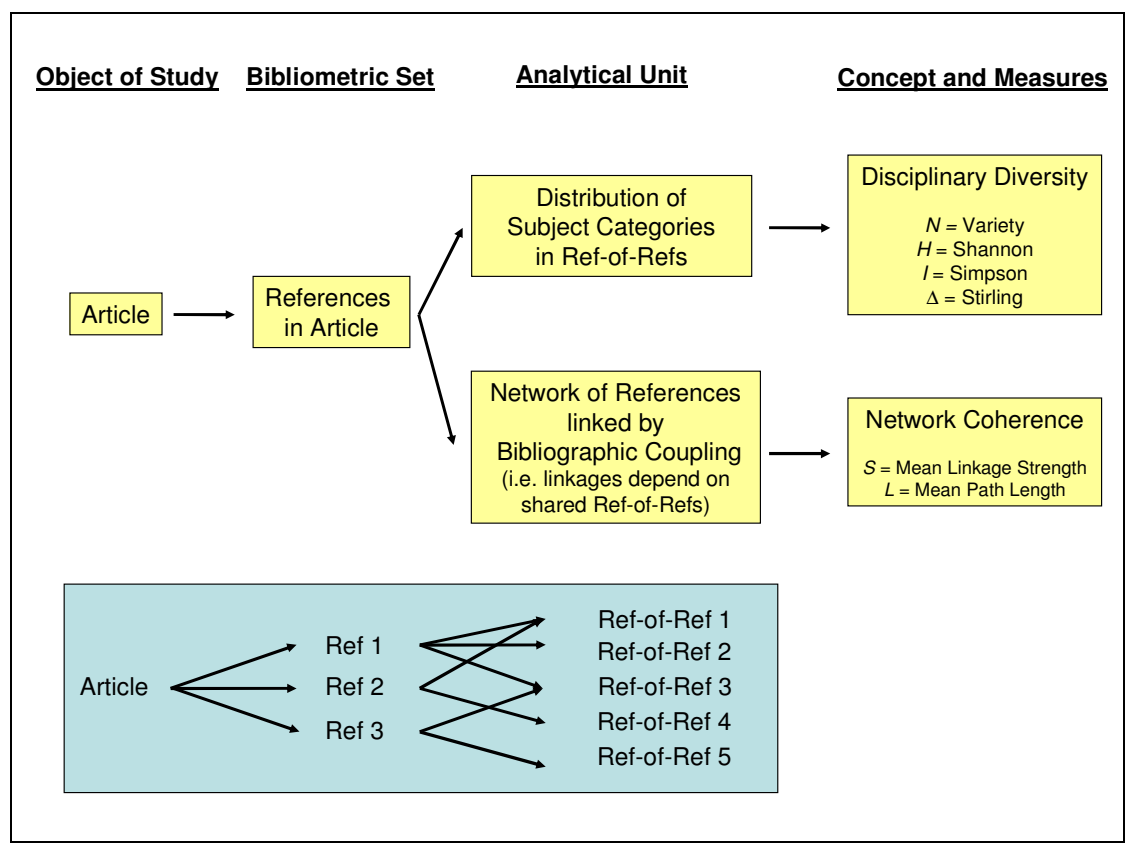

Figure 3. Scheme of operationalisation of disciplinary diversity and network coherence for one article.

### 3.2. Operationalisation of disciplinary diversity

The disciplinary diversity of an article was constructed from the distribution of ISI SCs in the references of references (*ref-of-refs* in Figure 3, and hereafter) of an article.[7] To compute this distribution, we constructed a frequency list of the journals in which the ref-of-refs were published, and converted it into a frequency list of ISI SCs using the SC attribution of each journal as given in the *Journal Citation Reports*. The mean for each article was 30 references (range 17 to 55), and 1,290 ref-of-refs (range 601 to 2,227). We cleaned the list for misnamed journals until at least 90% of the ref-of-refs in each list were attributable (average attribution rate: 95%).

The distribution of SCs in the ref-of-refs list allowed us to compute variety $N$ as the number of SCs that appeared at least once, and the Shannon $H$ and Simpson $I$ diversities (see Table 1). All indicators were normalised to a value between zero and 1.[8] In order to compute the Stirling $\Delta$ diversity, a similarity matrix $s_{ij}$ for the SCs must be constructed. To do so, we created a matrix of citation flows matrix between SCs, and then converted it into a Salton's cosine similarity matrix in the citing dimension. The $s_{ij}$ describes the similarity in the citing patterns for each pair of SCs in 2006, for the SCI set (175 SCs). A detailed description and analysis of this $s_{ij}$ SC-similarity matrix is provided elsewhere (Leydesdorff and Rafols, submitted). By combining the ref-of-refs SC proportions $p_i$ and similarities $s_{ij}$, we computed $\Delta$ as shown in Table 1. This particular operationalisation of the Stirling $\Delta$ diversity yields an indicator that is almost identical to Porter et al.'s (2007) Integration.

Using the $s_{ij}$ similarity matrix we constructed science maps in terms of SCs (Figure 4), similar to those reported in Moya-Anegón et al. (2004, 2007). The labels in Figure 4 describe clusters of similar SCs derived from factor analysis (Leydesdorff and Rafols, submitted). Following Klavans and Boyack (2008), we used the science map as a 'backbone' on which to overlay the distribution of SCs from each article, to provide an intuitive perspective of the position of its knowledge base in the scientific landscape (Scharnhorst, 1998).

---

[7] If the initial bibliometric set is large enough for statistical purposes, diversity can be computed directly from the SCs of the references.
[8] Simpson $I$ and Stirling $\Delta$, by definition, satisfy this condition. Variety $N$ and Shannon $H$ are normalised by dividing by their maximum values, $N_{max}$ and $ln(N_{max})$, respectively, with $N_{max}$ being the total number of ISI SCs.

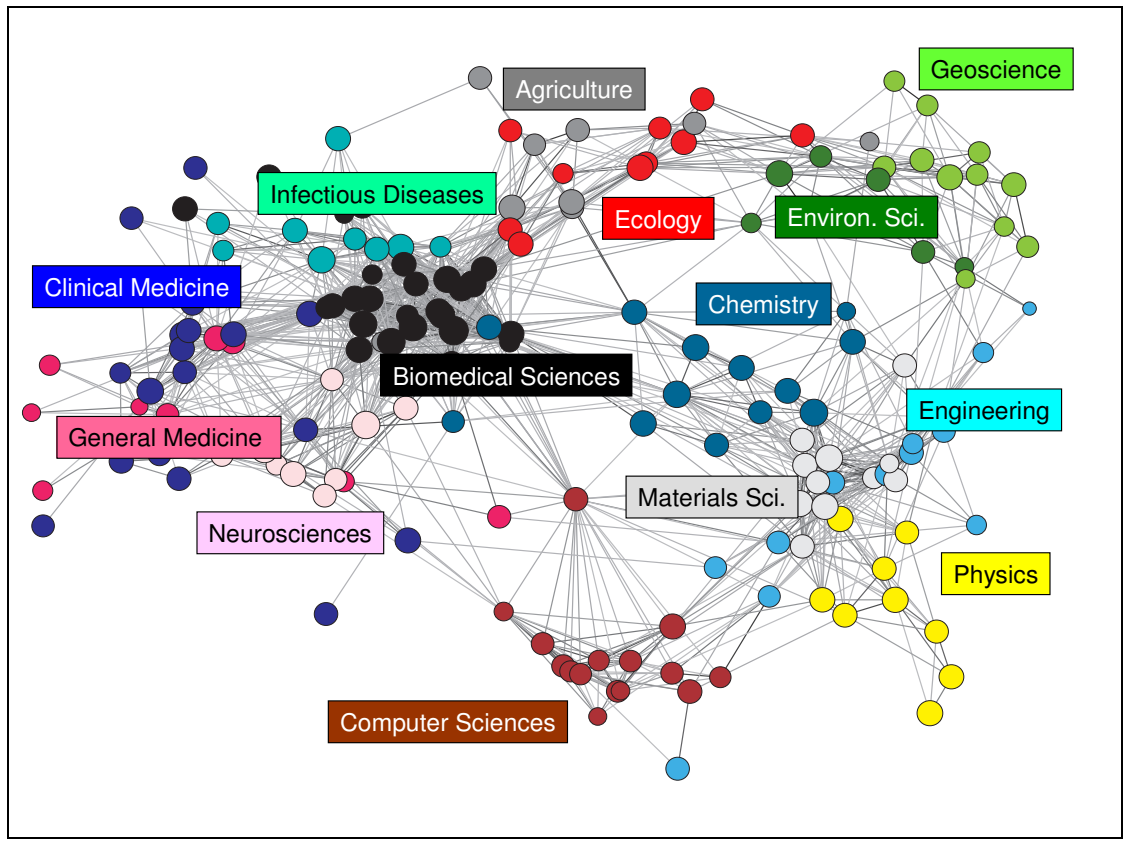

*Figure 4. Map of science for 2006 based on similarities in citing patterns between ISI Subject Categories. Based on Leydesdorff and Rafols (submitted).*

### 3.3. Operationalisation of network coherence

In order to operationalise network coherence for our bibliometric set, we chose first, a similarity metric between network elements (articles) in order to measure the strength of their linkages; second, an indicator of structural coherence of the network. Since the aim was to map the breadth of knowledge sources, similarity was measured in terms of bibliographic couplings[9] between articles (co-occurrences of references), and normalised using Salton's cosine (Ahlgren et al., 2003). Then, basic network measures were used as indicators for network coherence:

- **Mean linkage strength, *S*:** the mean of the bibliographic coupling matrix, excluding the diagonal - equivalent to network density in binary networks. In valued networks, it describes both realised links and intensity of similarities. By definition, *S* has a value between zero and 1.

---

[9] Although co-citation analysis is the most extended technique to measure similarities between publications, it is impractical for our purposes for two reasons: first, it cannot be for used for recently published papers, due to lack of citations; second, it reflects similarities in the audience, rather than in the knowledge sources.

- **Mean path length, *L*:** the path length between two articles is defined as the minimum number of links crossed to go from one article to the other over the network. Mean path length describes how 'spread' the network is; it is computed after binarising similarities.

These measures can be interpreted in terms of network centralities, which were introduced in bibliometrics to study research communities (Otte and Rousseau, 2002) and interdisciplinarity in journal sets (Leydesdorff, 2007). Mean linkage strength *S* is the mean degree centrality normalised by network size; mean path length *L* is equal to the mean of closeness centrality. More sophisticated measures of network 'compactness' or cohesion (Egghe and Rousseau, 2003) are not used in this study, but deserve further exploration within the conceptual framework proposed.

Given that network measures are generally highly size dependent, it is necessary to check the scale invariance of *S* and *L*. Since the bibliometric networks in these case studies are small (between 17-55 articles), we tested empirically the scale invariance of *S* in an independent sample of 1,275 articles related by research topic (kinesin). It was found that *S* was size independent and that the distribution of bibliographic couplings could be approximated to a log-normal distribution.[10] This result is in accordance with Havemann et al. (2007), who used a similar approach. Since *S* and *L* are highly correlated in our sample (Pearson = 0.95), *L*'s size dependence appears to be negligible as well in this study.

A possible drawback to bibliographic coupling is that it relates articles that share only one or two very general references, e.g. classical methodological handbooks, which do not necessarily inform about shared expertise. In order to minimise these spurious connections, we set a threshold of linkage strength to 0.05=1/20 when computing path length -as a result even in the smallest reference sets (20), at least two common references are needed for two papers to be linked.[11]

## 4. Case studies in molecular motors

The objective of this study is to assess the degree of interdisciplinarity of individual contributions in the specialty of molecular motors. As explained in the Data and Methods section, we build on previous investigations that carried

---

[10] Details of the scale invariance test are presented below.

| Network size | 10 | 51 | 255 | 637 | 1275 |
|---|---|---|---|---|---|
| Mean linkage strength | 0.022 | 0.023 | 0.024 | 0.025 | 0.024 |
| Standard deviation per network | 0.045 | 0.046 | 0.047 | 0.047 | 0.046 |
| Network realisations | 10 | 9 | 7 | 1 | 1 |
| Standard deviation over realisations | 0.007 | 0.004 | 0.002 | -- | -- |

From a network of 1,275 publications on kinesin research, random subnetworks of different sizes were extracted. Mean linkage strength and standard deviation were computed for each. For small networks, multiple realisations were carried out to minimise statistical fluctuations.

[11] In one case, Noji 1997, we had to set the threshold at 0.025 in order to keep the network connected.

out detailed case studies on interdisciplinary practices in five research projects (Rafols and Meyer, 2007; Rafols, 2007). It emerged from interviews that the techniques and concepts in all cases came from a variety of disciplines and that their combined use was crucial for the succes of the research. For example, in one case, newly developed fluorescence microscopy (biophysics) was combined with genetically engineered biomolecular motors (molecular biology) in order to trace the displacement of the motor at the nano-scale [12]. However, in spite of this shared interdisciplinarity, in some cases the projects were a continuation of a well-established research tradition and built on a narrow literature, while in others the projects brought together different research traditions and previously unrelated literatures. Can the indicators of disciplinary diversity and network coherence capture these differences?

We present the results of five case studies based on analysis of 12 articles. Table 3 shows the distribution of SCs in the ref-of-refs for each article. Given the well-documented innacuray of ISI SCs (Boyack et al., 2005), these proportions should be taken as indicative. *Biochemistry and Molecular Biology* is the dominant discipline, but there are also important contributions from *Cell Biology* and *Biophysics*. Records in *Multidisciplinary Sciences* journals constitute almost 25% of the total, possibly obscuring the actual distribution of references among the top SCs.[13] In particular, qualitative data from interviews suggested a larger presence of biophysics. After the four top SCs, the proportions are much smaller, and the distribution tails of some articles differ, e.g. Funatsu (1995) has a 'fatter' tail for the physical and chemical disciplines.

Table 3. Distribution of Subject Categories for each article.

| Subject Category (% ref-of-refs) | Fun 95 | Koj 97 | Ish 98 | Noj 97 | Yas 98 | Oka 99 | Kik 01 | Sak 99 | Bur 03 | Tom 00 | Tom 02 | Yil 04 | Cum Ave. |
|---|---|---|---|---|---|---|---|---|---|---|---|---|---|
| Biochem. & M. Biol. | 31.1 | 23.9 | 45.5 | 49.2 | 51.9 | 32.2 | 36.7 | 30.5 | 30.8 | 32.4 | 30.8 | 27.4 | 36.2 |
| Multidiscip. Sci. | 29.0 | 32.6 | 20.7 | 17.2 | 13.0 | 32.4 | 25.5 | 22.3 | 15.6 | 26.9 | 24.0 | 26.7 | 59.4 |
| Cell Biology | 10.4 | 19.2 | 6.6 | 8.8 | 8.9 | 21.2 | 21.0 | 29.2 | 37.5 | 26.3 | 30.7 | 28.7 | 80.4 |
| Biophysics | 9.8 | 7.9 | 12.5 | 14.2 | 15.8 | 6.0 | 6.6 | 6.0 | 4.5 | 5.6 | 4.3 | 5.6 | 88.5 |
| Physiology | 4.3 | 2.5 | 6.8 | 1.7 | 2.5 | 1.2 | 0.8 | 1.2 | 0.5 | 0.9 | 0.6 | 0.9 | 90.5 |
| Bio. Res. Meth. | 1.5 | 2.6 | 1.7 | 1.7 | 1.9 | 1.8 | 1.4 | 3.5 | 1.6 | 1.2 | 1.1 | 0.9 | 92.1 |
| Gen. & Heredity | 0.2 | 0.4 | 0.0 | 0.6 | 0.2 | 0.3 | 0.8 | 0.8 | 1.7 | 0.9 | 1.2 | 2.1 | 92.9 |
| Neurosciences | 0.2 | 0.4 | 0.1 | 0.1 | 0.1 | 1.5 | 0.6 | 0.4 | 0.5 | 1.1 | 3.1 | 1.9 | 93.7 |
| Biology | 0.6 | 0.2 | 1.1 | 0.8 | 1.5 | 0.3 | 0.7 | 1.0 | 1.3 | 0.4 | 0.3 | 0.0 | 94.5 |
| Crystallography | 0.0 | 0.1 | 0.5 | 0.2 | 0.1 | 0.6 | 3.0 | 0.0 | 0.0 | 0.5 | 0.2 | 0.3 | 95.1 |
| Develop. Biology | 0.2 | 0.1 | 0.0 | 0.0 | 0.0 | 0.3 | 0.7 | 0.2 | 1.1 | 1.1 | 1.5 | 1.3 | 95.7 |
| Chem., Analytical | 1.2 | 0.6 | 0.3 | 0.4 | 0.8 | 0.8 | 0.1 | 1.4 | 0.6 | 0.2 | 0.1 | 0.1 | 96.2 |
| Optics | 1.5 | 1.9 | 0.5 | 0.4 | 0.1 | 0.1 | 0.0 | 0.3 | 0.0 | 0.2 | 0.2 | 0.8 | 96.5 |
| Chem., Phys. | 2.6 | 0.1 | 0.4 | 0.4 | 0.4 | 0.1 | 0.0 | 0.1 | 0.2 | 0.0 | 0.0 | 0.5 | 96.8 |
| Chem., Multidisc. | 1.2 | 0.1 | 0.8 | 0.1 | 0.2 | 0.1 | 0.1 | 0.2 | 0.4 | 0.1 | 0.0 | 0.2 | 97.1 |
| Phys,A.Mol.&Chem. | 1.7 | 0.2 | 0.3 | 0.8 | 0.6 | 0.0 | 0.1 | 0.0 | 0.1 | 0.0 | 0.0 | 0.2 | 97.3 |
| Physics, Multidisc. | 1.1 | 1.1 | 0.3 | 0.2 | 0.1 | 0.3 | 0.0 | 0.0 | 0.0 | 0.3 | 0.2 | 0.5 | 97.6 |

---

[12] More qualitative insights are described in Rafols and Meyer (2007) and Rafols (2007).
[13] This might explain, in part, the large difference between the SC distribution of ref-of-refs in Table 4 and the distribution of references among four selected disciplines reported in Rafols and Meyer (2007).

Legend: Columns show the distribution of ref-of-refs for one article. Those in the same box belong to the same project. The last column is the cumulative percentage averaged over all cases.

### 4.1. Comparison between indicators

Table 4 presents the measures for disciplinary diversity and network coherence for each article; Table 5 presents the correlations between the different diversity and coherence measures. Figure 5 plots diversity $\Delta$ vs coherence $S$ for each article.

We compare, first, indicators, and, second, articles. Diversities $H$, $I$ and $\Delta$ were found to be correlated. Interestingly, the highest correlation was between Shannon $H$ and Stirling $\Delta$, although Stirling $\Delta$ and Simpson $I$ (rather than Shannon) have similar mathematical formulations. Since Shannon $H$ gives more weight to the small terms in its sum through its logarithmic factor, while Stirling $\Delta$ gives more weight to the combinations of disparate SCs, we believe that the high correlation between $H$ and $\Delta$ is due to the fact that many SCs with small proportions happen also to be distant from the core SCs.

Indicators of coherence, $S$ and $1/L$, were also highly correlated with one another, but not with the diversity measures. Hence, we are capturing two different aspects of the same bibliometric set. Variety $N$ was not correlated with any other measure, and it does not seem to be a good indicator of knowledge integration. Given this set of correlations, and in order to simplify the analysis, the discussion that follows is based on Stirling's $\Delta$ for diversity and mean linkage path $S$ for coherence.

Table 4. Measures of disciplinary diversity and network coherence

| Articles | Disciplinary Diversity | | | | Network Coherence | |
|---|---|---|---|---|---|---|
| | N | H | I | $\Delta$ | S | 1/L |
| Funatsu 1995 | 0.16 | 0.39 | 0.79 | 0.27 | 0.054 | 0.54 |
| Kojima 1997 | 0.20 | 0.38 | 0.79 | 0.24 | 0.074 | 0.70 |
| Ishijima 1998 | 0.22 | 0.34 | 0.72 | 0.18 | 0.042 | 0.53 |
| Noji 1997 | 0.18 | 0.32 | 0.70 | 0.15 | 0.024 | 0.43 |
| Yasuda 1998 | 0.19 | 0.31 | 0.68 | 0.14 | 0.039 | 0.54 |
| Okada 1999 | 0.14 | 0.32 | 0.74 | 0.15 | 0.107 | 0.73 |
| Kikkawa 2001 | 0.20 | 0.33 | 0.75 | 0.16 | 0.072 | 0.63 |
| Sakakibara 1999 | 0.15 | 0.34 | 0.77 | 0.16 | 0.029 | 0.47 |
| Burgess 2003 | 0.20 | 0.34 | 0.74 | 0.14 | 0.050 | 0.59 |
| Tomishige 2000 | 0.19 | 0.33 | 0.75 | 0.14 | 0.104 | 0.69 |
| Tomishige 2002 | 0.16 | 0.33 | 0.75 | 0.15 | 0.113 | 0.79 |
| Yildiz 2004 | 0.18 | 0.35 | 0.77 | 0.17 | 0.065 | 0.58 |
| Mean | 0.18 | 0.34 | 0.75 | 0.17 | 0.064 | 0.60 |
| Stand. Deviation | 0.02 | 0.02 | 0.03 | 0.04 | 0.030 | 0.11 |

Legend: $N$ = variety of disciplines, $H$ = Shannon, $I$ = Simpson, $\Delta$ = Stirling, $S$ = mean sinkage strength, $L$ = mean path length. Indicators are normalised to a value between zero and 1. Highest diversity and lowest coherence values are highlighted.

Table 5. Pearson correlations between diversity and coherence measures

| Pearson's Correlations | Disciplinary Diversity | | | | Network Coherence | |
|---|---|---|---|---|---|---|
| | N | H | I | Δ | S | 1/L |
| N | 1.00 | 0.04 | -0.20 | -0.01 | -0.23 | -0.10 |
| H | | 1.00 | 0.81 | 0.95 | -0.12 | -0.06 |
| I | | | 1.00 | 0.71 | 0.32 | 0.31 |
| Δ | | | | 1.00 | -0.10 | -0.06 |
| S | | | | | 1.00 | 0.95 |
| 1/L | | | | | | 1.00 |

Legend: $N$ = variety of disciplines, $H$ = Shannon, $I$ = Simpson, $\Delta$ = Stirling, $S$ = mean linkage strength, $L$ = mean path length. Highest correlations are highlighted.

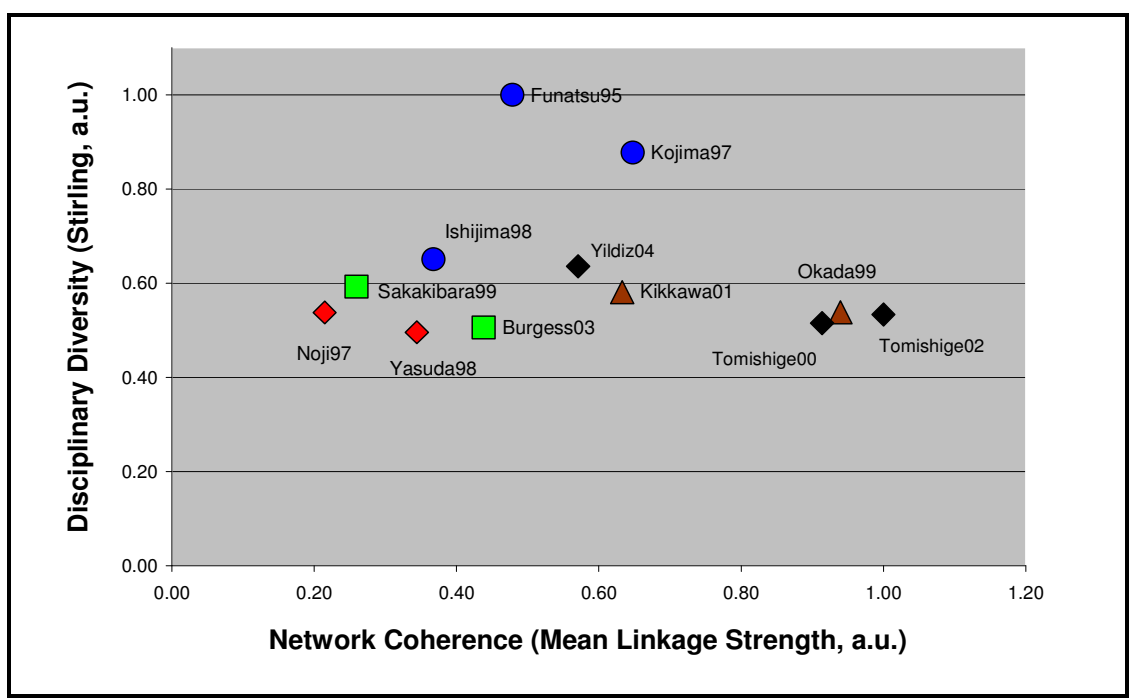

*Figure 5. Disciplinary diversity vs network coherence. Same shape and colour indicate same project. Data is presented in arbitrary units (a.u.) obtained by dividing a series by the largest value.*

### 4.2. Comparison among articles

Since we do not have benchmarks for diversity or coherence from other areas of science, we cannot investigate whether this field is (or is not) particularly diverse or coherent. However, this does not preclude comparison within our set. To do so, we combine the data presented in Table 4 and Figure 5, with visualisations of the SC distributions and network structures respectively in Figures 6 and 7.

Figure 6 visualises the relative contribution of SCs to an article over the backbone map of science, on the basis of the ref-of-refs distribution. Here the size of each SC node is arbitrarily set to a logarithmic factor of its SC proportion (i.e. $Area = \ln(1 + 1000 \cdot p_i)$) in order to facilitate visualisation of small SCs. The map shows that the most highly cited SCs are in the area of biomedical sciences and closely related to one another. There are a few contributions from nearby areas such as neuroscience, and a tail of contributions spanning from chemistry to some areas of physics.

In line with information gleaned from the interviews, the disciplinary distributions in the science map are very similar for all the articles. The exceptions are Funatsu 1995 and Kojima 1997, which have thicker tails – see Figure 6 and compare maps for Funatsu 1995 (top) with that of Noji 1997 (bottom). This is congruent with the indicators in Figure 5. Funatsu 1995 and Kojima 1997 are publications from a research group composed mainly of biophysicists, which made major contributions to the development of single molecule microscopy and manipulation. The distribution in Table 3 shows that the share of biophysics is not particularly high, but there are sizeable proportions of physics and chemistry related disciplines. Ex-post, it could be argued that the physics tail is consistent with the type of physics-based insights and techniques needed to develop single molecule microscopy and manipulation. Since physics and biological sciences have a large cognitive distance (see Figure 4), their interaction would have a larger weight in Stirling Δ.

However, ex-ante, based on the qualitative investigation, we did not assess Funatsu and Kojima's group to be any more interdisciplinary than the others on this axis, and we remain cautious in claiming higher disciplinary diversity for this group. For example, Yildiz 2004 is a publication that also developed single molecule microscopy based on biophysics, yet it does not present a physics related tail in its SC distribution. Hence, the two exception cases showing higher Δ, cast doubt on the reliability of the disciplinary diversity indicator. Our unit of analysis, individual publication, may be too small for the ISI SC categorisation, which is known to be coarse-grained; e.g. Boyack et al., 2005, report a more than 50% disagreement between journal-based clustering and SCs. We expect improved reliability from use of finer-grained and/or more accurate taxonomies, e.g. based on bottom-up large-scale mapping efforts (Boyack et al., 2005).

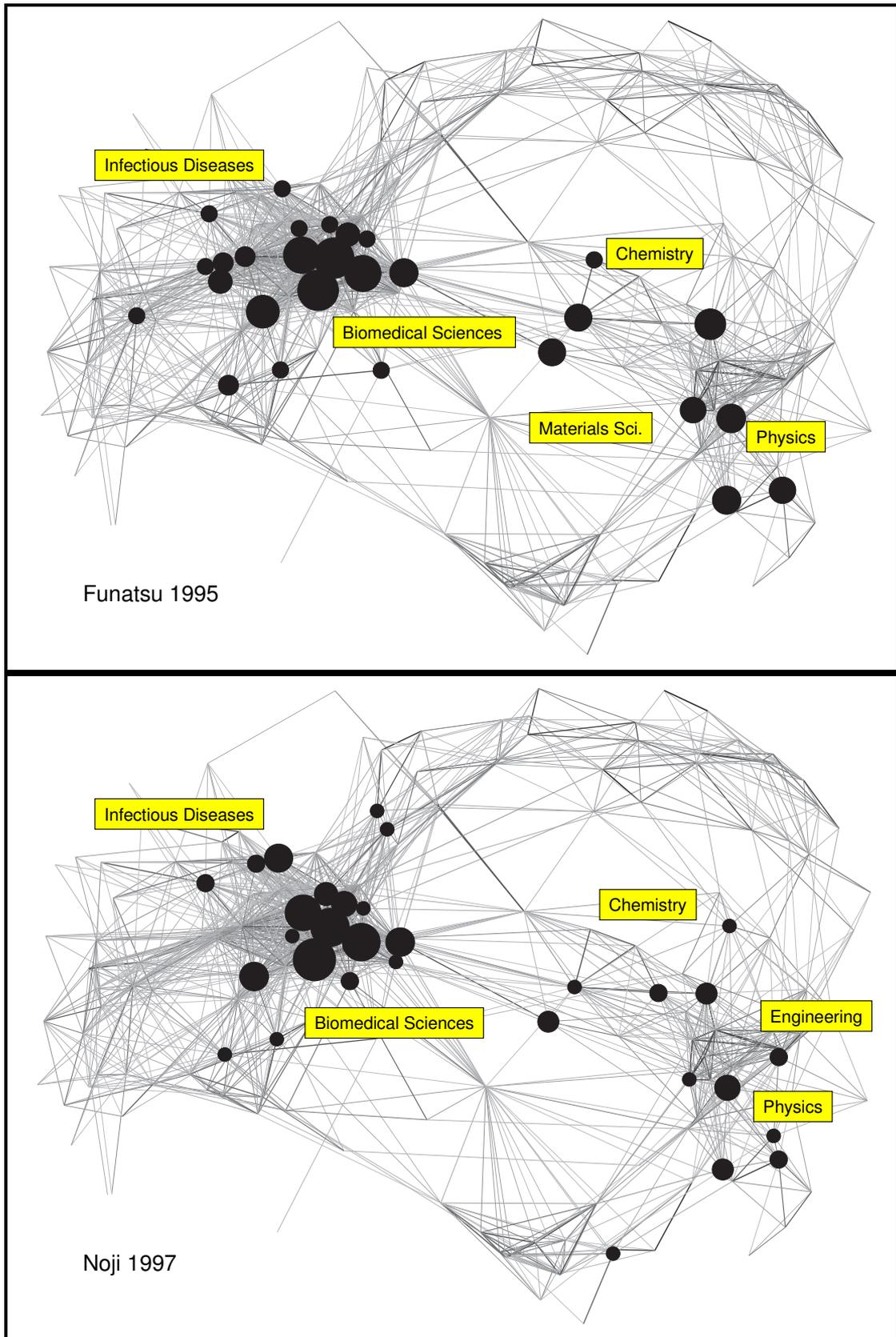

Figure 6. Distribution on the map of science of Subject Categories (SCs) of ref-of-refs in an article. The area of nodes is a logarithmic factor of a SC proportion in the ref-of-refs distribution, i.e. $\ln(1+1000p_i)$.

Network coherence values, on the other hand, vary widely among case studies, as shown in Figure 5. Figure 7 illustrates the differences in network structure associated with increasing values for coherence, for four articles. The first, Noji 1997, is an interesting case of convergence of two strands of research by two laboratories in one collaborative project spawned by a PhD student (Noji). Noji's lab was mainly based in biochemistry and worked on $F_1$-ATPase, a protein complex in the mitochondria studied by a research community focused on bioenergetics. The publication Noji 1997 was the result of a close collaboration with a biophysics laboratory specialised in linear molecular motors (myosin and kinesin). The network of Noji's references depicted in Figure 7, neatly illustrates the divide in the literature between the two research communities: on the right hand side, are publications on linear molecular motors; on the left, are publications on bioenergetics ($F_1$-ATPase). The only (weak) link between the two is due to a review with more than 311 references.[14] The low value for network coherence captures the fact that this article brought together distant bodies of knowledge, i.e. it would fall in the upper left quadrant in Figure 2.

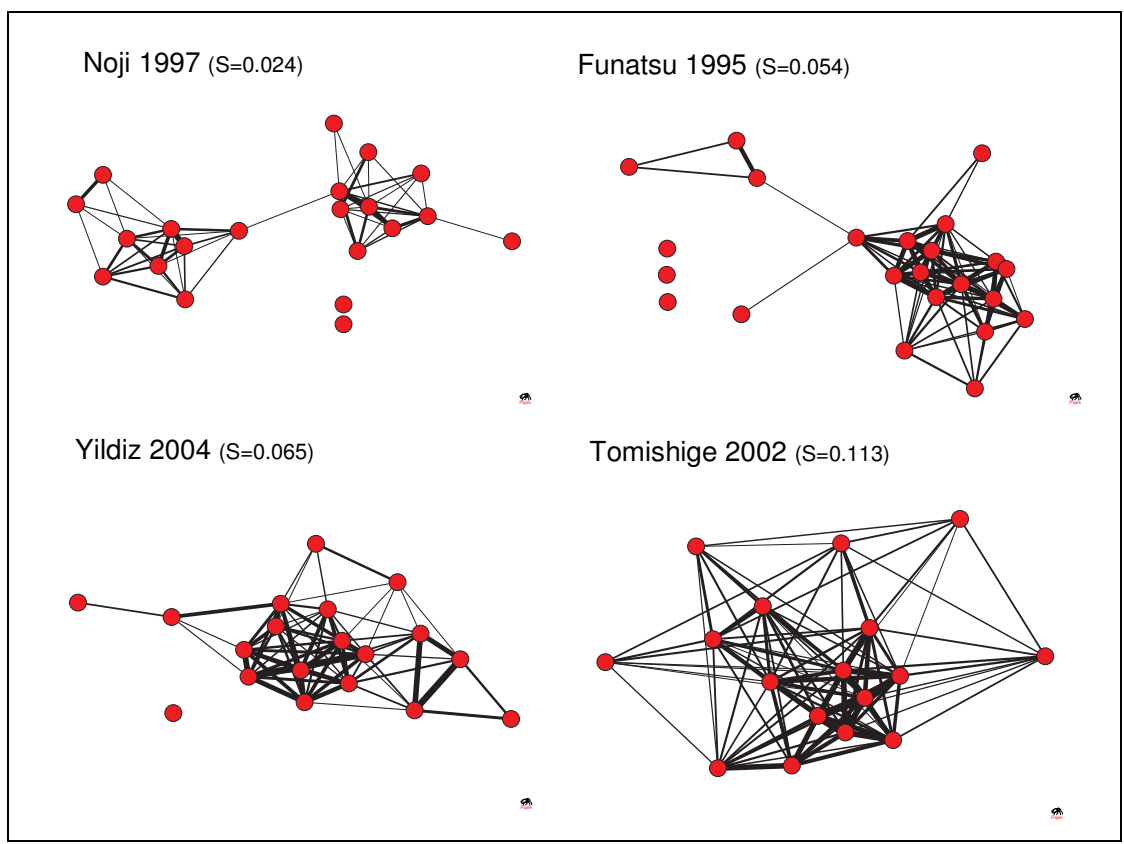

*Figure 7. Bibliographic coupling networks for the reference set of various articles. The figures are ordered from lower to higher network coherence S (from top left to bottom right); thicker lines indicate greater similarity.*

---

[14] The historical anecdote is that Paul D. Boyer, the author of this long review, was awarded the Nobel Prize precisely in 1997, thanks, in part, to the evidence provided by Noji and co-authors on his model of ATPase as a rotary motor.

Figure 8 illustrates the disciplinary mix of the Noji 1997 reference set by locating the SC where references were published. Given that Noji's project was a collaboration between a biophysics and a biochemistry laboratory, it could be expected that the bioenergetics cluster would publish mainly on biochemistry, and the molecular motors cluster on biophysics. However, both clusters have publications in biochemistry, biophysics and cell biology,[15] which suggests that bioenergetics research on its own, and molecular motors on its own scored high for disciplinary diversity prior to Noji's paper. In other words, this is a case of convergence of two bodies of knowledge that were already interdisciplinary.

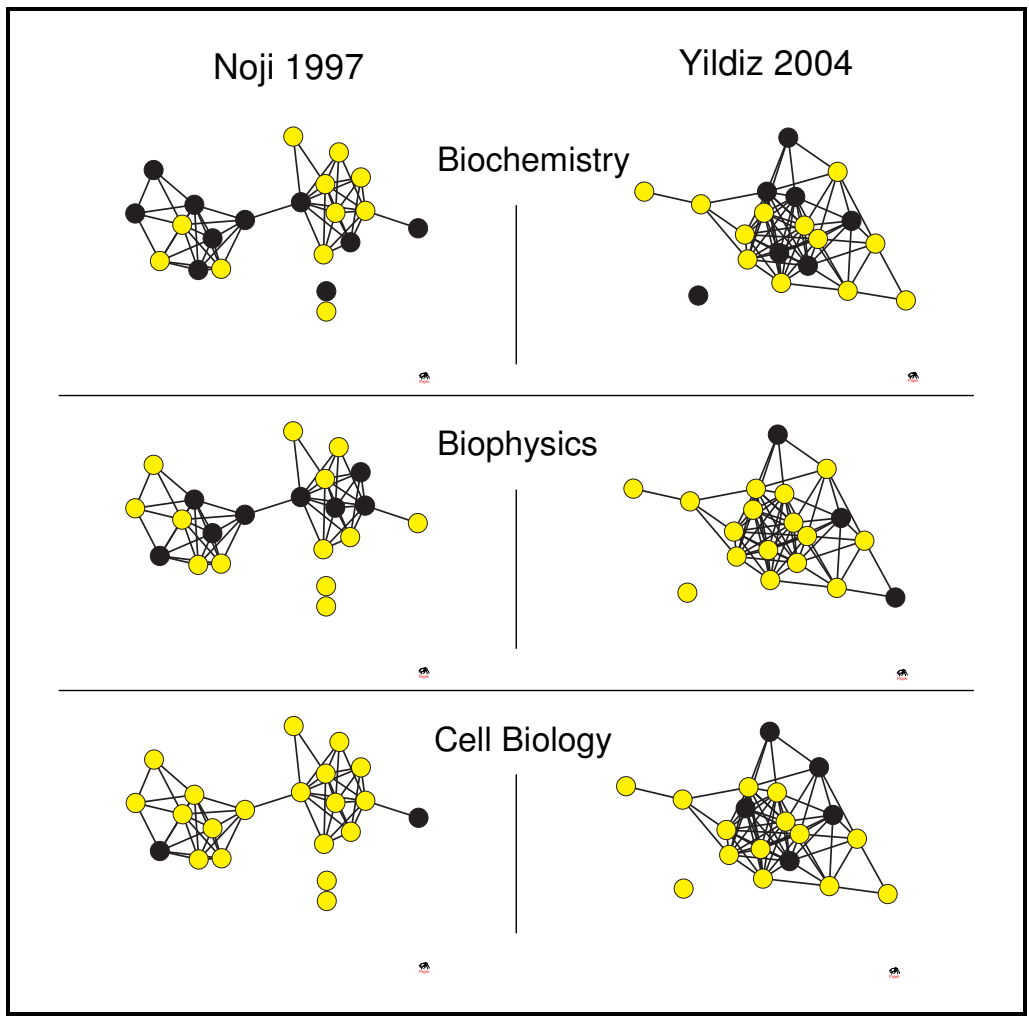

*Figure 8. Distribution of SCs for publications in bibliographic coupling networks. Black nodes indicate the papers published in a given ISI subject category.*

The second case of network coherence we examine is Yildiz 2004 (Figure 7 - bottom left). This article was also the result of a collaborative project, between a biophysics laboratory with expertise in fluorescent microscopy and Vale's laboratory, one of the leading molecular motors groups, which has an eclectic knowledge mix of cell biology and biophysics (including fluorescent microscopy expertise). Vale's lab contributed a genetically modified protein that they had engineered in a previous study, and Yildiz's lab contributed a new type of single molecule microscopy. Since both teams were working in the same specialty (molecular motors) and had some overlapping expertise in single molecule microscopy, they already shared a cognitive base. Hence, the references in their joint publication form a coherent cluster. Nevertheless, the cluster contains publications in biochemistry, cell biology and biophysics – the three main areas of molecular motors research, as shown in Figure 8 (right). Therefore, Yildiz 2004 appears to be a case of research within an already integrated (or specialised) interdisciplinary body of knowledge (upper right quadrant in Figure 2). Tomishige 2002 (a previous publication from Vale's lab, see Figure 7, bottom right) is an even more 'compact' example of an already specialised interdisciplinary publication – although, as explained, Vale's approach is integrative in the disciplinary sense: he draws on knowledge and recruits researchers from biophysics, cell biology and related fields.

---

[15] Two caveats apply to Figure 8: (i) on average 30% of the references were published in *Multidisciplinary Sciences* journals; (ii) about 25% of the references are published in journals that are attributed to at least two SCs (which is why the publication SCs cannot be presented in one unique network).

Finally, we have Funatsu 1995 (top right in Figure 8), which is an intermediate case between Noji 1997 and Yildiz 2004. This article reports a technical breakthrough in single molecule visualisation by a team mainly based in biophysics and well established within the research specialty of molecular motors. Hence, not surprisingly, molecular motors constitutes the main body of the literature in the dense cluster, in the lower left of Figure 8. However, the group also drew on its unique microscopy expertise, which extended beyond molecular motors. This 'external' expertise is exemplified by the three detached papers in the network which dealt exclusively with microscopy from a physical science perspective. Thus, Funatsu 1995 would be a case of acquisition of external supplementary knowledge from one literature (technical studies of microscopy) into the main cluster of molecular motors research. This limited integration effort would locate this publication in the upper middle part of Figure 2.

These examples suggest that bibliographic coupling networks and the network coherence indicators derived from them, provide a suitable tool for examination of the processes of knowledge integration at local level. The limitation, as discussed in Section 2.6, is that these micro perspectives cannot assess how different are the bodies of knowledge integrated in the larger context of science.

From the dual perspective of diversity and coherence, the case studies investigated provide empirical evidence that publications with similar levels of disciplinary diversity could be at very different stages of knowledge integration: Noji 1997 would be an example of an incipient interdisciplinary knowledge integration process (upper left in Figure 2), Tomishige 2002, an example of interdisciplinary specialised research (upper right) and Funatsu 1995, an intermediate case. Hence, molecular motors research appears to be spread over the upper part of Figure 2 (relatively high disciplinary diversity),[16] covering the left and right quadrants.

## 5. Conclusions

### 5.1. Summary of analytical framework and results

In this article, we proposed a novel conceptual framework to investigate interdisciplinary processes in the wider sense of knowledge integration. The framework is based on the concepts of diversity and coherence, borrowed respectively from ecology and network analysis (Table 2 and Figure 2), and already used implicitly in previous bibliometric studies on interdisciplinarity (e.g. Morillo et al., 2003). Diversity was used to capture the disciplinary heterogeneity of our bibliometric set as seen through the filter of predefined categories, i.e. taking a top-down perspective in order to locate the set on the global map of science (Figure 6). Coherence was used to apprehend the intensity of similarity relations within the bibliometric set, i.e. using a bottom-up approach to reveal the structural consistency and cognitive articulation of the publications network (Figure 7).

Disciplinary diversity indicators were constructed from diversity indices (Shannon $H$ and Simpson $I$) and a recently developed indicator (Stirling $\Delta$, parameterised as Porter's Integration), which takes account of the similarities between SCs (Stirling 1998, 2007; Porter et al., 2007). ISI SCs were used as disciplinary categories. Network coherence was operationalised in terms of the network measures Mean linkage strength and mean path length, in bibliographic coupling networks (see Havemann et al., 2007 for a similar approach). These indicators were applied to the reference set of publications in a bionanoscience research specialty, biomolecular motors, for which we had detailed information from interviews (Rafols and Meyer, 2007; Rafols, 2007).

First, we found that the indicators for disciplinary diversity and network coherence were not correlated (Table 4), thus providing 'orthogonal' perspectives of the knowledge integration process. Among diversity indicators, Shannon $H$ and Stirling $\Delta$ made more salient the contributions of small or disparate categories.

Second, disciplinary diversity took similar values for most of the publications examined, in line with our previous qualitative investigations (Table 4 and Figure 5). However, there are grounds to cast some doubt on the reliability of this indicator, given its unexpected high values for two publications (out of 12) and low proportion of biophysics in the SC distributions. Since there is a trade-off between accuracy and simplicity of a taxonomy, it is possible that the unit of analysis (the article) in this study is too small for the coarse-grained description of science provided by ISI SCs. Comparative studies using different disciplinary taxonomies (e.g. provided by other bibliometric databases or categories derived from large-scale clustering) would be needed to ascertain the scope of reliable application.

---

[16] This is an inference from the qualitative interviews. Without quantitative benchmarks from other areas of science, the position of the case studies on the disciplinary diversity axis cannot be determined.

Third, we found that measures for network coherence could discriminate among articles according to their different degrees of knowledge integration at micro level. For example, the case of lowest network coherence (Noji 1997), was the result of a collaboration between two laboratories based on two different bodies of knowledge (bioenergetics vs linear molecular motors). On the contrary, those cases with high network coherence (such as Tomishige 2002) were based on only one research tradition (molecular motors), although they still relied on several disciplines. We believe that the discrimination between these two different phases of knowledge integration (seminal integration vs specialisation in already integrated areas, depicted in Figures 2 and 7), is important in emergent fields such as nanotechnology and systems biology, in order to distinguish pioneering integrative efforts from less risky rides on 'interdisciplinary bandwagons'.

The operationalisation of network coherence in terms of mean linkage strength of bibliographic coupling appeared to work well, both for our small sets and in larger studies reported by Havemann et al. (2007). Moreover, it has the advantage of simplicity. However, there is scope for exploring more sophisticated measures of network coherence (e.g. Egghe and Rousseau, 2003), and more nuanced cognitive similarities between publications (e.g. including co-word analysis, as in Braam et al., 1991).

Fourth, the visualisations of diversity (through the overlay of disciplinary proportions on the map of science, Figure 6), and of coherence (by means of the bibliographic coupling network, Figure 7), proved more valuable than expected. Although initially developed to support the indicators, the maps and networks provide a richer and subtler representation of the different aspects of diversity (variety, balance and similarity) and coherence (linkage strength, density, clustering), which characterise the knowledge integration process.

Fifth, the differences in network coherence observed for publications with similar disciplinary diversity, support the view that interdisciplinarity is an inadequate term or a misnomer (Klein, 2000, p.3; Gläser, 2006). In focusing on knowledge integration, adopting a bottom-up approach and looking at emergent structures, we encounter fuzzy and overlapping bodies of knowledge, as illustrated in Figure 7, that do not conform to established categories. In our view, the crucial dynamics of knowledge integration lies in the interactions between these local bodies of knowledge. The use of macro (disciplinary) categories only provides information on the position of these local bodies on the science map.

### 5.2. Future research and possible applications

This study has developed a conceptual framework and methodology for capturing knowledge integration in research, which we applied to small case studies. How robust and generalizable is this approach? We think that this pilot study should be extended in the following directions for the method to be fully validated. First, benchmarks with other areas of science need to be established in order to gauge the range of high/low values on the diversity and coherence axes; second, investigations employing larger bibliometric sets are needed to check scalability; and third, studies using different taxonomies should test the sensitivity of disciplinary diversity to differences in categorisations. The approach proposed could easily be adopted and adapted, at least for small and medium sized bibliometric sets (e.g. $10^4$ records), given that it is based on simple indicators and standard similarity measures. Most of these can be computed using very simple bibliometric tools (freeware) and publicly available data.[17]

Regarding the degree of general applicability, we believe the approach could be directly utilised, with little modification, for a number of science policy issues, including:

**1) Evaluation of interdisciplinary programmes**: Porter et al. (2007) report the use of the Integration indicator (equivalent to our development of Stirling $\Delta$) for evaluation of interdisciplinary performance of researchers involved in the National Academies Keck Future Initiative, on the basis of their publication records. The inclusion of an indicator for network coherence may add an orthogonal perspective;

**2) Emergence and diffusion of research topics:** We have conducted preliminary studies on quantum dots (with A.L. Porter) and kinesin research, investigating diffusion and knowledge integration patterns from their appearance in a narrow field of science to their spread into broader research areas. Here the aim is to identify the key integrative research communities in the diffusion/translation process. We think that this use of diversity and coherence indicators can be valuable in comparative studies of emergent and 'hyped' fields such as nanotechnology, where claims of novelty and interdisciplinarity are rife, but not always substantiated.

**3) Evaluation of diversity in science:** Concerns have been expressed that diversity in the science system might be declining as a result of the increasing dependence of funding on performance evaluation (Schmidt et al., 2006). Diversity and coherence indicators offer the possibility to address this issue through longitudinal or national comparative studies, as in Havemann et al. (2007).

Finally, we would point to the benefits of basing our approach on a general conceptual framework. First, in using a general framework, the concepts underlying the current indicators of interdisciplinarity are rendered more transparent. This, in turn, facilitates discussion of their inevitable biases, and adaptation to social and policy needs. Second, the generality of the formulation allows its application and cross-fertilisation among distinct research areas. Thus, we expect insights and enriching perspectives of the diversity-coherence framework from ongoing investigations on technological diversification in firms, biodiversity, energy portfolio and similar system approaches (Nesta and Saviotti, 2005; Stirling, 2007).

---

[17] The only processed input needed is the SC similarity matrix used to create the science map and compute Stirling $\Delta$. This is available as a Pajek input file (Leydesdorff and Rafols, submitted).: http://users.fmg.uva.nl/lleydesdorff/map06/data.xls

## Acknowledgements

This research was supported by an EU postdoctoral Marie-Curie Fellowship to IR, and the Daiwa Anglo-Japanese Foundation. We benefited from discussions with J. Gläser, L. Leydesdorff, F. Morillo, A.L. Porter, and SPRU colleagues S. Katz, R. Kempener, W.E. Steinmueller and A. Stirling.